\documentclass[12pt]{article}
\usepackage{amsmath,bbm,amssymb}
\usepackage{amsthm}
\usepackage{comment}
\usepackage{graphicx, graphics}
\usepackage{textcomp}
\usepackage{amsfonts}

\usepackage{psfrag,pst-node,subfigure,rotating, amsmath, bbm, amsthm, amssymb, amsthm, setspace, picture, epsfig, amsfonts, upgreek}

\usepackage{multirow}

\usepackage{arydshln}
\usepackage{algorithm}
\usepackage{algorithmic}

\usepackage{arydshln}

\newtheorem{assumption}{Assumption}

\newtheorem{theorem}{Theorem}
\newtheorem{lemma}{Lemma}

\def\m{\mathbf}\def\v{\boldsymbol}\def\s{\mathcal}

\begin{document}
\begin{center}
{\Large\bf A unified framework of principal component analysis and factor analysis}
\\[2mm]     Shifeng Xiong$^{1,2}$\footnote{Corresponding author, Email: xiong@amss.ac.cn}
{\footnotesize\\ 1. NCMIS, KLSC, Academy of Mathematics and Systems Science, Chinese Academy of Sciences\\ Beijing, China
\\[1mm] 2. School of Mathematical Sciences, University of Chinese Academy of Sciences\\ Beijing, China}
\end{center}

\vspace{1cm} \noindent{\bf Abstract}\quad Principal component analysis and factor analysis are fundamental multivariate analysis methods. In this paper a unified framework to connect them is introduced. Under a general latent variable model, we present matrix optimization problems from the viewpoint of loss function minimization, and show that the two methods can be viewed as solutions to the optimization problems with specific loss functions. Specifically, principal component analysis can be derived from a broad class of loss functions including the $\ell_2$ norm, while factor analysis corresponds to a modified $\ell_0$ norm problem. Related problems are discussed, including algorithms, penalized maximum likelihood estimation under the latent variable model, and a principal component factor model. These results can lead to new tools of data analysis and research topics.

\vspace{1cm} \noindent{{\bf KEY WORDS:} Dimensionality reduction; Latent variable model; Low-rank approximation; Matrix optimization; Probabilistic principal component analysis.}


\newpage

\section{Introduction}\label{sec:intro}

Principal component analysis (PCA) and factor analysis (FA) can be found in any textbook of multivariate statistical analysis; see e.g. Anderson (1984). The two methods both have long histories of over one hundred years. As basic dimensionality reduction methods, PCA and FA reduce the dimension of the original random vector via eigenvectors of the covariance matrix and interpretable factors, respectively. Their variants and applications have been extremely studied in the literature (Jolliffe 2002; Bartholomew, Knott, and Moustaki 2011).

Generally speaking, PCA is model free, but FA is based on the factor model. There are some connections between PCA and FA. Tipping and Bishop (1999) used a special factor model to define a probabilistic PCA method. PCA is often used as the start point in the iterations of computing maximum likelihood estimates under the factor model (Jolliffe 2002). Overall, PCA and FA are viewed as two relatively separate topics. There is no work on more general and closer connections between the two fundamental methods. In this paper we introduce a framework to unify them within a general latent variable model. We show that they can be viewed as solutions that minimize specific loss functions under this model. Specifically, PCA can be derived from a broad class of loss functions including the $\ell_2$ norm, while FA corresponds to a modified $\ell_0$ norm loss. Besides, other main contributions of this paper include:
\begin{description}
\item[](a) We show that PCA can be derived from a penalized maximum likelihood estimation problem. When treating the penalty as a Bayesian prior, we obtain a Bayesian version of PCA, which can be viewed as a new probabilistic PCA model.
\item[](b) We propose a penalized least squares method for FA, which performs well in terms of dimensionality reduction under unidentifiable factor models. The corresponding iterative algorithm is also given.
\item[](c) We present a combination of PCA loss and FA loss, and construct the corresponding path between PCA and FA.
\item[](d) We introduce a principal component factor model that links FA to PCA more closely.
\end{description}To the best of our knowledge, the above findings have not appeared in the literature. It is expected that these findings cannot only help us better understand PCA and FA, but also lead to new tools of data analysis and research topics.

The rest of this paper is organized as follows. Section \ref{sec:review} briefly reviews PCA and FA. Section \ref{sec:mo} introduces the general latent variable model and related optimization problems. Sections \ref{sec:fpca} and \ref{sec:ffa} present the objective functions corresponding to PCA and FA, respectively. Section \ref{sec:cb} discusses combinations of PCA and FA. Section \ref{sec:num} provides numerical illustrations. We conclude the paper with some discussion in Section \ref{sec:diss}. All proofs are given in the Appendix.

\section{Review of principal component analysis and factor analysis}\label{sec:review}

In this section we present a brief review of principal component analysis (PCA) and factor analysis (FA). Some notation and definitions are needed. For a vector or matrix $\m{a}$, let $\m{a}'$ denote its transpose. For two symmetric matrices $\m{A}$ and $\m{B}$ of the same size, write $\m{A}\leqslant$ $(\geqslant)$ $\m{B}$ if $\m{B}-\m{A}$ ($\m{A}-\m{B}$) is nonnegative definite, and write $\m{A}<$ $(>)$ $\m{B}$ if $\m{B}-\m{A}$ ($\m{A}-\m{B}$) is positive definite. Let $\m{0}$ denote the zero matrix.

Consider a random $p$-vector $\v{X}$ whose covariance matrix $\mathrm{Cov}(\v{X})=\m{\Sigma}>\m{0}$. Without loss of generality, let its expectation $\mathrm{E}(\v{X})=\v{0}$. PCA of $\v{X}$ is based on the eigenvalue decomposition of $\m{\Sigma}$, denoted by $\m{\Sigma}=\m{R}\m{\Lambda}\m{R}'$, where $\m{R}=(\v{r}_1,\ldots,\v{r}_p)$ is an orthogonal matrix and $\m{\Lambda}$ is a diagonal matrix $\mathrm{diag}(\lambda_1,\ldots,\lambda_p)$ with $\lambda_1\geqslant\cdots\geqslant\lambda_p>0$. The $k$th principal component of $\v{X}$ is $\v{r}_k'\v{X}$, whose variance is $\lambda_k$.
Partition $\m{R}$ as $\m{R}=(\m{R}_1\ \m{R}_2)$, where $\m{R}_1\in\mathbb{R}^{p\times q}$. Let $\m{\Lambda}_1=\mathrm{diag}(\lambda_1,\ldots,\lambda_q)$. The cumulative proportion of the first $q$ principal components $\m{R}_1'\v{X}$ to $\v{X}$ is defined as $(\lambda_1+\cdots+\lambda_q)/(\lambda_1+\cdots+\lambda_p)=\mathrm{trace}\left(\m{\Lambda}_1\right)/\mathrm{trace}\left(\m{\Sigma}\right)
=\mathrm{trace}\left(\mathrm{Cov}(\m{R}_1'\v{X})\right)/\mathrm{trace}\left(\mathrm{Cov}(\v{X})\right)$. If the cumulative proportion is sufficiently large, say, at least $85\%$, then we can say that $\v{X}$ is well represented by the first $q$ principal components. Hence PCA is a linear dimensionality reduction technique. In practice, PCA can be implemented through decomposing an estimate of $\m{\Sigma}$, say, the sample covariance matrix, with observations of $\v{X}$.

FA of $\v{X}$ is based on the following factor model, \begin{equation}\label{fm}\v{X}=\m{A}\v{U}+\v{\varepsilon},\end{equation}where the factor loading matrix $\m{A}$ is an $p\times q$ constant matrix of full column rank, $1\leqslant q<p$, the common factors $\v{U}\sim N_q(\v{0},\m{I}_q)$ is independent of $\v{\varepsilon}\sim N_p(\v{0},\m{V})$, $\m{I}_q$ is the identity matrix of order $q$, and $\m{V}=\mathrm{diag}(v_1,\ldots,v_p)$ with $v_i\geqslant0,\ i=1,\ldots,p$. We have $\mathrm{Cov}(\v{X})=\m{\Sigma}=\m{A}\m{A}'+\m{V}>\m{0}$. The main purpose of FA is to estimate $\m{A}$, or $\m{A}\m{A}'$, which yields a low-dimensional representation of $\v{X}$. Note that the factor model \eqref{fm} can be unidentifiable. Usually we need to make the following assumption.

\begin{assumption}\label{ass:id} Model \eqref{fm} is identifiable (Shapiro 1985), i.e., there does not exist a nonnegative definite diagonal matrix $\tilde{\m{V}}$, different from $\m{V}$, such that $\m{\Sigma}-\tilde{\m{V}}\geqslant\m{0}$ and $\mathrm{rank}(\m{\Sigma}-\tilde{\m{V}})\leqslant\mathrm{rank}(\m{\Sigma}-\m{V})$.\end{assumption}

Next we provide some comparisons between PCA and FA. A major distinction between them is that there is a definite model underlying FA, while PCA does not require a model. Tipping and Bishop (1999) proposed a probabilistic model for PCA, which is actually a special case of \eqref{fm} with $v_1=\cdots=v_p$; see also Roweis (1997). This model cannot cover the broad range of applications of PCA. An advantage of FA is its interpretation because the common factors $\v{U}$ in \eqref{fm} usually have real meanings, and this is the main reason why FA is so popular in various fields. In addition, PCA is often used in computations of FA (Jolliffe 2002; Bartholomew, Knott, and Moustaki 2011). For detailed review of PCA and FA, we refer readers to several monographes and review papers, e.g., Bartholomew, Knott, and Moustaki (2011), Jolliffe and Cadima (2016), and Ghojogh, Crowley, and Karray et al. (2023).

\section{A latent variable model and related matrix optimization problems}\label{sec:mo}

Consider the following latent variable model that generalizes \eqref{fm}, \begin{equation}\label{lvm}\v{X}=\m{A}\v{U}+\v{\varepsilon}.\end{equation}
Like in \eqref{fm}, $\m{A}$ is an $p\times q$ constant matrix of full column rank, $q<p$, $\v{U}\sim N_q(\v{0},\m{I}_q)$ is an unobserved latent vectors independent of $\v{\varepsilon}\sim N_p(\v{0},\m{V}),\ \m{V}\geqslant\m{0}$. Generally, $\v{V}$ is not restricted to be diagonal. However, we sometimes restrict it to be a diagonal or other special matrix according to the problem. Write $\m{T}=\m{A}\m{A}'\geqslant\m{0}$, and thus $\mathrm{rank}(\m{T})=q$. By \eqref{lvm}, $\mathrm{Cov}(\v{X})=\m{\Sigma}=\m{A}\m{A}'+\m{V}=\m{T}+\m{V}$. Unlike the factor model, any $\m{\Sigma}$ can be written as this form. Clearly the model \eqref{lvm} is unidentifiable, and we can specify $\m{T}$ and $\m{V}$ through minimizing various loss functions. In the following, $\m{\Sigma}$ also denotes an estimate of the population covariance matrix when only observations of $\v{X}$ are available.

We consider two classes of methods for specifying $\m{T}$ and $\m{V}$. One is to first specify $\m{T}$ by solving the following matrix optimization problem,
\begin{equation}\label{mo}
	\begin{aligned}
		&\min_{\m{T}}\ F_1(\m{T}),\\
		&{\text{s.t.}}\ \m{0}\leqslant\m{T}\leqslant\m{\Sigma},\ \mathrm{rank}(\m{T})=q,
	\end{aligned}
\end{equation}where $F_1$ is an objective function defined on the set of all nonnegative definite matrices. Consequently, we specify $\m{V}$ as $\m{V}^*=\m{\Sigma}-\m{T}^*$ (or other strategies, since $\m{V}$ is less important), where $\m{T}^*$ is the solution to \eqref{mo}.
The other is to simultaneously specify $\m{T}$ and $\m{V}$ without the requirement $\m{T}+\m{V}=\m{\Sigma}$ through solving the two-matrix optimization problem,
\begin{equation}\label{mo2}
	\begin{aligned}
		&\min_{\m{T},\m{V}}\ F_2(\m{T},\m{V}),\\
		&{\text{s.t.}}\ \m{T}\geqslant\m{0},\ \m{V}\geqslant\m{0},\ \mathrm{rank}(\m{T})=q,
	\end{aligned}
\end{equation}where $F_2$ is an objective function.

The two-matrix problem can yield a one-matrix problem by taking
\begin{equation}\label{2to1}F_1(\m{T})=F_2(\m{T},\m{V}^*(\m{T}))\end{equation} in \eqref{mo}, where $\m{V}^*(\m{T})$ is the minimum of $F_2(\m{T},\m{V})$ for given $\m{T}$. Conversely, one-matrix problems induce two-matrix problems under the following assumption. Let $h$ be a function defined on the set of all nonnegative definite matrices.

\begin{assumption}\label{ass:h} For all $\m{S}\geqslant\m{0}$, $h(\m{S})\geqslant h(\m{\Sigma})$.\end{assumption}

\begin{theorem}\label{th:F2} Let $\m{T}^*$ be the solution to \eqref{mo}. If Assumption \ref{ass:h} holds, then $(\m{T}^*, \m{\Sigma}-\m{T}^*)$ is the solution to \eqref{mo2} with $F_2(\m{T},\m{V})=h(\m{T}+\m{V})+F_1(\m{T})$.
\end{theorem}

In the following two sections we will show that PCA and FA can be derived by \eqref{mo} or \eqref{mo2} with specified loss/objective functions.

\section{Objective functions corresponding to PCA}\label{sec:fpca}

\subsection{One-matrix optimization}\label{subsec:pcaomo}

In the viewpoint of dimensionality reduction, it is required that the distribution of $\v{X}$ can be well approximated by that of $\m{A}\v{U}$, i.e., $\m{T}$ should be close to $\m{\Sigma}$. In other words, we need to find a low-rank approximation of $\m{\Sigma}$. This inspires us to consider the objective function $F_1(\m{T})=f(\m{\Sigma}-\m{T})$, where $f$ is a function defined on the set of all nonnegative definite matrices that evaluates the distance between a nonnegative definite matrix and the zero matrix. To induce PCA, we present the following assumptions on $f$.
\begin{assumption}\label{ass:f0} For all $\m{S}\geqslant\m{0}$, $f(\m{S})\geqslant0$ and $f(\m{S})=0$ if and only if $\m{S}=\m{0}$.\end{assumption}

\begin{assumption}\label{ass:f} For all $\m{S}\geqslant\m{0}$ and orthogonal matrix $\m{R}$ of order $p$, $f(\m{R}\m{S}\m{R}')=f(\m{S})$.\end{assumption}

\begin{assumption}\label{ass:fd} For all $\m{S}=(s_{ij})_{i,j=1,\ldots,p}\geqslant\m{0}$, $f(\m{S})\geqslant f\left(\mathrm{diag}(s_{11},\ldots,s_{pp})\right)$. \end{assumption}

\begin{assumption}\label{ass:fm} The function $g(\v{x})=f\left(\mathrm{diag}(\v{x})\right),\ \v{x}\in[0,+\infty)^p$, is nondecreasing in each variable.\end{assumption}

Consider the following $f$: the trace $f(\m{S})=\mathrm{trace}(\m{S})$, the $\ell_2$ norm (also called spectral norm) $f(\m{S})=\|\m{S}\|_2$ which is the largest eigenvalue of $\m{S}$, and the Frobenius norm $f(\m{S})=\|\m{S}\|_F=(\mathrm{trace}(\m{S}^2))^{1/2}$.

\begin{lemma}\label{lemma:afd} The $\ell_2$ norm $f(\m{S})=\|\m{S}\|_2$ satisfies Assumption \ref{ass:fd}.
\end{lemma}

By Lemma \ref{lemma:afd} and simple algebra, $f={\mathrm{trace}}(\m{S}),\ \|\m{S}\|_2$, and $\|\m{S}\|_F$ satisfy Assumptions \ref{ass:f0}-\ref{ass:fm}. Actually, Assumptions \ref{ass:f}-\ref{ass:fm} are technical, designed only for the following proofs. Besides the above $f$'s and their combinations such as $f(\m{S})=\|\m{S}\|_2+\|\m{S}\|_F+{\mathrm{trace}}(\m{S})$, it is hard to find other $f$ to satisfy the assumptions.

\begin{theorem}\label{th:pca} Under Assumptions \ref{ass:f0}-\ref{ass:fm}, the solution to \eqref{mo} with $F_1(\m{T})=f(\m{\Sigma}-\m{T})$ is $\m{T}^*=\m{R}_1\m{\Lambda}_1\m{R}'_1$.
\end{theorem}

Theorem \ref{th:pca} indicates that PCA can be derived from the optimization problem \eqref{mo} with a broad class of objective functions. Taking $f(\m{S})={\mathrm{trace}}(\m{S})$ corresponds to the common concept of PCA, variance maximization. Besides, other objective functions that do not satisfy the above assumptions can yield PCA. For example, let $f(\m{S})={\mathrm{rank}}(\m{S})$, which does not satisfy Assumption \ref{ass:fd}. Note that ${\mathrm{rank}}(\m{\Sigma}-\m{T})\geqslant{\mathrm{rank}}(\m{\Sigma})-{\mathrm{rank}}(\m{T})=p-q={\mathrm{rank}}(\m{\Sigma}-\m{T}^*)$. The solution to \eqref{mo} with $f(\m{S})={\mathrm{rank}}(\m{S})$ is $\m{T}^*=\m{R}_1\m{\Lambda}_1\m{R}'_1$.

Interestingly, if we expand the domain of $f$ to the set of all symmetric matrices, then PCA also corresponds to the solution to the optimization problem \eqref{mo} without the constraint $\m{T}\leqslant\m{\Sigma}$, i.e.,
\begin{equation}\label{fmo}
	\begin{aligned}
		&\min_{\m{T}}\ f(\m{\Sigma}-\m{T}),\\
		&{\text{s.t.}}\ \m{T}\geqslant\m{0},\ \mathrm{rank}(\m{T})=q,
	\end{aligned}
\end{equation}for $f$ satisfying the further assumption.
\begin{assumption}\label{ass:sym} For all symmetric $\m{S}$, $f(-\m{S})=f(\m{S})$.\end{assumption}
The following theorem can be proven similarly to Theorem \ref{th:pca}.
\begin{theorem}\label{th:pcae} Under Assumptions \ref{ass:f0}-\ref{ass:sym}, the solution to \eqref{fmo} is $\m{T}^*=\m{R}_1\m{\Lambda}_1\m{R}'_1$.
\end{theorem}

It is clear that $f(\m{S})=\|\m{S}\|_2$ and $\|\m{S}\|_F$ satisfy Assumption \ref{ass:sym} while ${\mathrm{trace}}(\m{S})$ does not. The special case with $f(\m{S})=\|\m{S}\|_F$ of Theorem \ref{th:pcae} was proven by Srebro and Jaakkola (2003) with matrix derivatives.

\subsection{Penalized maximum likelihood estimation}\label{subsec:mle}

In this subsection we derive PCA by the likelihood principle. The negative log-likelihood (up to a constant) under \eqref{lvm} is
\begin{equation}\label{lh}
	l(\m{T},\m{V})=\log(|\m{T}+\m{V}|)+{\mathrm{trace}}\left((\m{T}+\m{V})^{-1}\m{\Sigma}\right),
\end{equation}and thus we can adopt $l$ as the objective function in the two-matrix optimization problem \eqref{mo2}. Clearly, this problem is not well defined since any $(\m{T},\m{V})$ satisfying $\m{T}+\m{V}=\m{\Sigma}$ is a solution. To derive PCA, we add a penalty and let
\begin{equation}\label{pmle}F_2(\m{T},\m{V})=l(\m{T},\m{V})+\lambda\,f(\m{\Sigma}-\m{T}),
\end{equation}where $\lambda>0$ is a pre-specified parameter.

Taking $h(\m{S})=\log(|\m{S}|)+{\mathrm{trace}}\left(\m{S}^{-1}\m{\Sigma}\right)$ in Theorem \ref{th:F2}, by Theorem \ref{th:pca}, we can get
\begin{theorem}\label{th:pmle} Under Assumptions \ref{ass:f0}-\ref{ass:fm}, for all $\lambda$, the solution to \eqref{mo2} with $F_2$ in \eqref{pmle} is $(\m{T}^*, \m{V}^*)=(\m{R}_1\m{\Lambda}_1\m{R}'_1, \m{R}_2\m{\Lambda}_2\m{R}'_2)$.
\end{theorem}
Theorem \ref{th:pmle} indicates that PCA can be viewed as a penalized maximum likelihood estimate. We can also view PCA as a Bayesian posterior mode under the following priors
\begin{equation}\label{prior}(\m{T}, \m{V})\propto\exp\left[-\lambda\,f(\m{\Sigma}-\m{T})\right]\cdot I(\m{T}\geqslant\m{0},\ \m{V}\geqslant\m{0},\ \mathrm{rank}(\m{T})=q),\end{equation}where $I$ denotes the indicator function. As a result we can obtain a Bayesian framework of PCA that allows uncertainty quantification for PCA. Unlike probabilistic PCA in Tipping and Bishop (1999), this probabilistic PCA framework accommodates to all covariance matrices.

Note that the priors in \eqref{prior} may be improper. We can also consider another type of penalties. For example, with a tuning parameter $\lambda>0$, let
\begin{equation}\label{pmle2}F_2(\m{T},\m{V})=l(\m{T},\m{V})+\lambda\,\|\m{V}\|_2^2,
\end{equation} which corresponds to the priors,
$$(\m{T}, \m{V})\propto\exp\left[-\lambda\,\|\m{V}\|_2^2\right]\cdot I(\m{T}\geqslant\m{0},\ \m{V}\geqslant\m{0},\ \mathrm{rank}(\m{T})=q).$$

\begin{theorem}\label{th:pmle2} For some $\lambda$, the solution to \eqref{mo2} with $F_2$ in \eqref{pmle2} is $(\m{T}^*, \m{V}^*)=(\m{R}_1\m{\Lambda}_1\m{R}'_1, \m{R}_2\m{\Lambda}_2\m{R}'_2)$.
\end{theorem}

\section{Objective functions corresponding to FA}\label{sec:ffa}

\subsection{One-matrix optimization}\label{subsec:omo}

Since $\m{V}=\m{\Sigma}-\m{T}$ is a diagonal matrix in the factor model \eqref{fm}, we first consider the $\ell_0$ norm that can yield sparsity.
For $\m{S}\geqslant\m{0}$, define its modified $\ell_0$ norm as \begin{equation}\label{f0}
	\tilde{f}_0(\m{S})=\left\lbrace \begin{aligned}
		&\|\m{S}\|_0, \quad\, |\m{S}|>0\\
		&p, \quad\quad\ \ |\m{S}|=0,\ \m{S}\neq\m{0},\\
        &0, \quad\quad\quad \m{S}=\m{0},
	\end{aligned}\right.
\end{equation}where $\|\m{S}\|_0$ and $|\m{S}|$ denote the number of nonzero elements in $\m{S}$ and the determinant of $\m{S}$, respectively. We have that $\tilde{f}_0$ satisfies Assumption \ref{ass:f0}, but does not satisfy Assumptions  \ref{ass:f}-\ref{ass:fm}.

\begin{theorem}\label{th:fa} Under model \eqref{fm}, $\m{T}^*=\m{A}\m{A}'$ is a solution to \eqref{mo} with $F_1(\m{S})=\tilde{f}_0(\m{\Sigma}-\m{S})$.
\end{theorem}

\begin{theorem}\label{th:fau} Under model \eqref{fm}, if Assumption \ref{ass:id} holds, then \eqref{mo} with $F_1(\m{S})=\tilde{f}_0(\m{\Sigma}-\m{S})$ has the unique solution $\m{T}^*=\m{A}\m{A}'$.
\end{theorem}

Besides the above $F_1$ constructed by the modified $\ell_0$ norm, many other objective functions can induce FA. For instance, let
\begin{equation}\label{fp}F_1(\m{S})=f_\tau(\m{\Sigma}-\m{S})\quad\text{and}\quad f_\tau(\m{S})=\sum_{i<j}|s_{ij}|^\tau,\quad\tau\geqslant 0
\end{equation}or\begin{equation}\label{fd}F_1(\m{S})=f_{\mathrm{d}}(\m{\Sigma}-\m{S})\quad\text{and}\quad f_{\mathrm{d}}(\m{S})=f(\m{S})-f({\mathrm{diag}}(s_{11},\ldots,s_{pp})),
\end{equation}where $f$ satisfies Assumption \ref{ass:fd}. Theorems \ref{th:fa} and \ref{th:fau} hold by replacing $\tilde{f}_0$ with $f_\tau$ or $f_{\mathrm{d}}$. By \eqref{2to1} and \eqref{lh}, the maximum likelihood estimates of $\m{T}$ can be derived from the objective function
\begin{equation}\label{lhfa}
	F_1(\m{S})=l(\m{S},\m{V}^*(\m{S})),
\end{equation}where $l$ is defined in \eqref{lh}, and $\m{V}^*(\m{S})$ is the solution that minimizes $l(\m{S},\m{V})$ under the constraint that $\m{V}\geqslant\m{0}$ is diagonal for given $\m{S}$.

\renewcommand{\algorithmicrequire}{\textbf{Inputs:}}
\renewcommand{\algorithmicensure}{\textbf{Steps:}}
\floatname{algorithm}{Algorithm}
\begin{algorithm}[t]
\caption{\label{ag:US}\quad The unidirectional search algorithm for solving \eqref{mo}}
\begin{algorithmic}[1]
\REQUIRE ~~\\  $\m{\Sigma},\ q,\ s,\ \varepsilon$. \ENSURE ~~\\\STATE \textbf{Initialization:} Select $\m{A}^{(0)}=(a_{ij}^{(0)})_{i=1,\ldots,p,\ j=1,\ldots,q}\in\mathbb{R}^{p\times q}$.
\\\STATE \textbf{Iteration:} For $k=0,1,\ldots,$\\\hspace{3mm} For $i=1,\ldots,pq$,\\\hspace{6mm} Change $a_{ij}^{(k)}$ in $\m{A}^{(k)}$ to $a_{ij}^{(k)}+s$, and let $\m{A}_+^{(ij)}$ denote the new matrix;
\\\hspace{6mm} Change $a_{ij}^{(k)}$ in $\m{A}^{(k)}$ to $a_{ij}^{(k)}-s$, and let $\m{A}_-^{(ij)}$ denote the new matrix.
\\\hspace{3mm} Find $\s{A}=\left\{\m{A}\in\left\{\m{A}_+^{(ij)},\ \m{A}_-^{(ij)}\right\}_{i=1,\ldots,p,\ j=1,\ldots,q}:\ F_1\left(\m{A}\m{A}'\right)<F_1\left(\m{A}^{(k)}\m{A}^{(k)'}\right),\ \m{A}\m{A}'\leqslant\m{\Sigma}\right\}$.
\\\hspace{3mm} If $\s{A}=\emptyset$, let $s\leftarrow s/2$;
\\\hspace{3mm} Else let $\m{A}^{(k+1)}=\arg\min_{\m{A}\in\s{A}}F_1(\m{A}\m{A}')$, and let $k\leftarrow k+1$. \\ If $s<\varepsilon$, then stop the iterations, and output $\m{T}^*=\m{A}^{(k)}\m{A}^{(k)'}$.
\end{algorithmic}
\end{algorithm}

Even when the factor model \eqref{fm} does not hold, we can solve \eqref{mo} with the objective functions in \eqref{f0}-\eqref{lhfa}. The corresponding solution can be viewed as an approximate FA approach for $\v{X}$, which extends FA to more general covariance structures such as the graphical factor models (Giudici and Stanghelline 2001; Grzebyk, Wild, and Chouani\`{e}re 2004).

The matrix optimization problem \eqref{mo} with the objective functions corresponding to FA are usually hard to solve. Here we provide an unidirectional search algorithm (Kolda, Lewis, and Torczon 2003) for solving \eqref{mo} with general objective functions; see Algorithm \ref{ag:US}. This is a direct search algorithm that does not require an explicit representation of the gradient of the objective function. The initial $\m{A}^{(0)}$ can be given by PCA. Actually Algorithm \ref{ag:US} is applicable to other objective functions including those corresponding to PCA. Note that Algorithm \ref{ag:US} may be computationally expensive, especially when $p$ is large. We need to develop more effective strategies in our future work.

\subsection{Two-matrix optimization}\label{subsec:tmo}

The most popular method for estimating the parameters of the factor model \eqref{fm} is maximum likelihood estimate (Anderson 1984). It is a two-matrix optimization with the likelihood objective function in \eqref{lh} and an additional constraint that $\m{V}$ is diagonal. Here we focus on another important estimation method, the least squares estimation (Bartholomew, Knott, and Moustaki 2011), which solves the two-matrix optimization problem,
\begin{equation}\label{mofa}
	\begin{aligned}
		&\min_{\m{T},\m{V}}\ \left\|\m{\Sigma}-(\m{T}+\m{V})\right\|_F^2,\\
		&{\text{s.t.}}\ \m{T}\geqslant\m{0},\ \mathrm{rank}(\m{T})=q,\ \m{V}\geqslant\m{0}\ \text{is diagonal}.
	\end{aligned}
\end{equation}This method does not require the normal assumption.

With matrix derivatives, iterative formulas for solving \eqref{mofa} can be derived (Bartholomew, Knott, and Moustaki 2011). We first show that these formulas can also be given by a coordinate descent algorithm (Tseng 2001). Specifically, given $\m{V}$, \eqref{mofa} reduces to the PCA problem \eqref{fmo}; given $\m{T}$, the solution is trivial by taking diagonal elements. Compared with existing algorithms, this coordinate descent algorithm is easy to understand and implement. Furthermore, from the coordinate descent viewpoint, the iterative formulas possess the monotonicity property that the objective value is non-increasing in each iteration.

\renewcommand{\algorithmicrequire}{\textbf{Input}:}
\renewcommand{\algorithmicensure}{\textbf{Steps}:}
\floatname{algorithm}{Algorithm}
\begin{algorithm}[t]
\caption{\label{ag:cd}\quad The coordinate descent algorithm for solving \eqref{pmlse}}
\begin{algorithmic}[1]
\REQUIRE ~~\\ $\m{\Sigma},\ q,\ \lambda,\ \varepsilon$. \ENSURE ~~\\\STATE \textbf{Initialization:} Select $\m{V}^{(0)}\leqslant\m{\Sigma}$. \STATE \textbf{Iteration:} For each $k=0,1,\ldots$,
\\\hspace{2mm} Compute $\m{T}^{(k+1)}=\bar{\m{R}}_1\bar{\m{\Lambda}}_1\bar{\m{R}}'_1$, where $\bar{\m{R}}_1$ and $\bar{\m{\Lambda}}_1$ correspond to the first $q$ eigenvectors and eigenvalues of $\m{\Sigma}-\m{V}^{(k)}$, respectively.
\\\hspace{2mm} Compute $\m{V}^{(k+1)}=\mathrm{diag}(\m{\Sigma}-\m{T}^{(k+1)})/(1+\lambda)$, where $\mathrm{diag}(\m{\Sigma}-\m{T}^{(k+1)})$ denotes the diagonal matrix whose diagonal is the same as $\m{\Sigma}-\m{T}^{(k+1)}$.
\\\hspace{2mm} If $\m{V}^{(k+1)}\leqslant\m{\Sigma}$ does not hold, then stop the iterations, and output $(\m{T}^*, \m{V}^*)=(\m{T}^{(k)},\m{V}^{(k)})$.
\\\hspace{2mm} Else if $\|\m{\Sigma}-\m{T}^{(k)}-\m{V}^{(k)}\|_F-\|\m{\Sigma}-\m{T}^{(k+1)}-\m{V}^{(k+1)}\|_F<\varepsilon$, then stop the iterations, and output $(\m{T}^*, \m{V}^*)=(\m{T}^{(k+1)},\m{V}^{(k+1)})$.
\\\hspace{2mm} Otherwise, let $k\leftarrow k+1$.
\end{algorithmic}
\end{algorithm}

When Assumption \ref{ass:id} does not hold, there are more than one pairs of $(\m{T},\m{V})$ satisfying the factor model \eqref{fm}. Note that $\m{V}$ should be close to the zero matrix from the viewpoint of dimensionality reduction. We next propose a penalized least squares estimation approach to produce the pair in which $\m{V}$ is  closest to the zero matrix. With a tuning parameter $\lambda\geqslant0$, the approach solves
\begin{equation}\label{pmlse}
	\begin{aligned}
		&\min_{\m{T},\m{V}}\ \left\|\m{\Sigma}-(\m{T}+\m{V})\right\|_F^2+\lambda\|\m{V}\|_F^2,\\
		&{\text{s.t.}}\ \m{T}\geqslant\m{0},\ \mathrm{rank}(\m{T})=q,\ \m{V}\geqslant\m{0}\ \text{is diagonal}.
	\end{aligned}
\end{equation} Suppose that $\m{\Sigma}$ is a $\sqrt{n}-$consistent estimate of the true covariance matrix, where $n$ denotes the sample size. If $\lambda=\lambda_n\to0,\ n\lambda_n\to+\infty$, then the solution to \eqref{pmlse} converges to the pair corresponding to the smallest value of $\|\m{V}\|_F$. Recall that $\|\m{\cdot}\|_F$ is an objective function to derive PCA. The penalized least squares approach can be viewed as a combination of PCA and FA. The coordinate descent algorithm can also be used to solve \eqref{pmlse}, and we present the detailed steps in Algorithm \ref{ag:cd}.


Like in Section \ref{subsec:mle}, we can derive penalized maximum estimation for FA and the corresponding Bayesian approach. Note that in these methods, $\m{V}$ is not restricted to be a diagonal matrix. Therefore,
we actually get a Bayesian version of the approximate FA approach for $\v{X}$, which can quantify the difference between the underlying model and a factor model.

\section{Combinations of PCA and FA}\label{sec:cb}
\subsection{Path optimization between PCA and FA}\label{subsec:path}

In this subsection we present a path optimization problem to connect PCA and FA. Consider the one-matrix optimization problem \eqref{mo}. Let $F_{1,\mathrm{PCA}}$ and $F_{1,\mathrm{FA}}$ denote two objective functions corresponding to PCA and FA, respectively. For $w\in[0,1]$, let $\m{T}_w^*$ denote the solution to
\begin{equation}\label{pcafa}
	\begin{aligned}
		&\min_{\m{T}}\ wF_{1,\mathrm{PCA}}(\m{T})+(1-w)F_{1,\mathrm{FA}}(\m{T}),\\
		&{\text{s.t.}}\ \m{0}\leqslant\m{T}\leqslant\m{\Sigma},\ \mathrm{rank}(\m{T})=q.
	\end{aligned}
\end{equation}Therefore $\{\m{T}_w^*\}_{w\in[0,1]}$ can be viewed as a path between PCA and FA. From the path we can find how the direction of dimensionality reduction varies from PCA to FA. This may help us better understand the difference between PCA and FA. The problem \eqref{pcafa} can be computed by Algorithm \ref{ag:US}. A simple choice of the objective functions in \eqref{pcafa} is $(F_{1,\mathrm{PCA}}(\m{T}),F_{1,\mathrm{FA}}(\m{T}))=(\|\m{\Sigma}-\m{T}\|_F^2,f_2(\m{\Sigma}-\m{T}))$, where $f_2$ is defined in \eqref{fp}.

Besides, \eqref{pcafa} provides a method to fit the factor model from the relatively simple PCA. For a sequence $1=w_0>w_1>\cdots<w_m=0$, we use PCA as the start point to compute the solution $\m{T}_{w_1}^*$, and then use $\m{T}_{w_1}^*$ as the start point to compute the solution $\m{T}_{w_2}^*$, and so on. Finally, $\m{T}_{w_m}^*$ obtained by this way is the FA solution. Note that the optimization problem with the FA objective function may have many local solutions. The solution from the path optimization is relatively close to PCA, which reaches some consistency between PCA and FA.

We can also consider two-matrix optimization for combining PCA and FA. The penalized least squares estimation approach in \eqref{pmlse} is an example. Similar to \eqref{pmlse}, \eqref{pcafa} can yield approximate estimates corresponding to the smallest PCA loss for unidentifiable factor models by taking $w\to0$.

\subsection{A principal component factor model}\label{subsec:pcfm}

As two dimensionality reduction methods, PCA and FA are sometimes required to have identical directions of dimensionality reduction (Tipping and Bishop 1999). In this subsection we introduce a special factor model, called principal component factor model, which adds a constraint \begin{equation}\label{con}\mathcal{L}(\m{A})=\mathcal{L}(\m{R}_1)\end{equation} to the original factor model \eqref{fm}, where $\mathcal{L}(\m{A})$ denotes the column space of $\m{A}$. Tipping and Bishop (1999)'s probabilistic PCA is a special case of the principal component factor model. 

We now discuss estimation for $\m{T}$ and $\m{V}$ under \eqref{con}. Given an objective function $G(\m{T},\m{V})$, we estimate them by solving the optimization problem,
\begin{equation}\label{pcfme}
	\begin{aligned}
		&\min_{\m{T},\m{V}}\ G(\m{T},\m{V}),\\
		&{\text{s.t.}}\ \m{T}=\mathcal{R}_1[\m{T}+\m{V}]\m{\Gamma}\mathcal{R}_1[\m{T}+\m{V}]',\ \m{\Gamma}\geqslant\m{0}\ \text{is diagonal},\m{V}\geqslant\m{0}\ \text{is diagonal},
	\end{aligned}
\end{equation}where $\mathcal{R}_1[\m{T}+\m{V}]$ denotes the $p\times q$ matrix corresponding to the first $q$ principal components of $\m{T}+\m{V}$. Note that the variables in $\m{\Gamma}$, which only appear in the constraint, are also needed to optimize.
The objective functions in \eqref{pcfme} corresponding to maximum likelihood estimation and least squares estimation for the principal component factor model can be found to in \eqref{lh} and \eqref{mofa}, respectively.

We also use the idea of coordinate descent to solve \eqref{pcfme}. Let $\v{\gamma}=(\gamma_1,\ldots,\gamma_q)'$ and $\v{v}=(v_1,\ldots,v_p)'$ be the diagonals of $\m{\Gamma}$ and $\m{V}$, respectively. Note that if $\mathcal{R}_1[\m{T}+\m{V}]=\m{R}_1$ is given, then \eqref{pcfme} reduces to a $(p+q)$-dimensional optimization problem
\begin{equation}\label{pcfm}
	\begin{aligned}
		&\min_{\v{\gamma},\v{v}}\ G(\m{R}_1{\mathrm{diag}}(\v{\gamma})\m{R}_1',{\mathrm{diag}}(\v{v})),\\
		&{\text{s.t.}}\ \gamma_1,\ldots,\gamma_q,v_1,\ldots,v_p\geqslant0.
	\end{aligned}
\end{equation} Therefore, we have the following iterative strategy. Given $\m{T}^{(k)}, \m{V}^{(k)}$, compute $\m{R}_1^{(k)}=\mathcal{R}_1[\m{T}^{(k)}+\m{V}^{(k)}]$; use \eqref{pcfm} to get $\m{\Gamma}^{(k)}, \m{V}^{(k+1)}$; compute $\m{T}^{(k+1)}=\m{R}_1^{(k)}\m{\Gamma}^{(k)}\m{R}_1^{(k)'}$.

For the objective function in least squares estimation, \eqref{pcfm} becomes the standard quadratic program problem,
\begin{equation}\label{qp}
	\begin{aligned}
		&\min_{\v{z}}\ \v{z}'\m{H}\v{z}-2\v{b}'\v{z},\\
		&{\text{s.t.}}\ z_i\geqslant{0},\ i=1,\ldots,p+q,
	\end{aligned}
\end{equation} where $\v{z}=(\v{\gamma}',\v{v}')'=(z_1,\ldots,z_{p+q})'$, $\m{H}=\begin{pmatrix} q\v{1}_q\v{1}_q' & \m{R}_{1}'\odot\m{R}_{1}' \\\m{R}_{1}\odot\m{R}_{1} & \m{I}_p \end{pmatrix}$, \\$\v{b}=\left(\sum_{i=1}^p\sum_{j=1}^p\sigma_{ij}\v{s}_{ij}',\,\sigma_{11},\ldots,\sigma_{pp}\right)'$, $\sigma_{ij}$ is the $(i,j)$-element of the current $\m{\Sigma}=\m{T}+\m{V}$, $\v{s}_{ij}=\v{r}^{(i)}\odot\v{r}^{(j)}$, $\v{r}^{(i)}$ is the $i$th row vector of $\m{R}_1$, and $\odot$ denotes element-wise multiplication. The solution to \eqref{qp} can be used as initial points in computing the maximum likelihood estimates.

The above principal component factor model can be extend to cover general factor models. Let $\m{R}_{(i)}$ denote the $p\times i$ sub-matrix of $\m{R}$ corresponding to the first $i$ eigenvectors, $1\leqslant i\leqslant p$. Consider the factor model \eqref{fm} with a factor loading matrix $\m{A}$. We define the principal component index of this model as the smallest integer $q*$ such that $\mathcal{L}(\m{A})\subset\mathcal{L}(\m{R}_{(q*)})$. Therefore, any factor model, with a principal component index at most $p$, is connected to PCA. In addition, for an unidentifiable factor model, we may improve its identifiability by reducing its principal component index.

\section{Numerical examples}\label{sec:num}

\subsection{Penalized least squares estimation}\label{subsec:ufm}

\begin{figure}[t]\begin{center}
\scalebox{0.7}[0.7]{\includegraphics{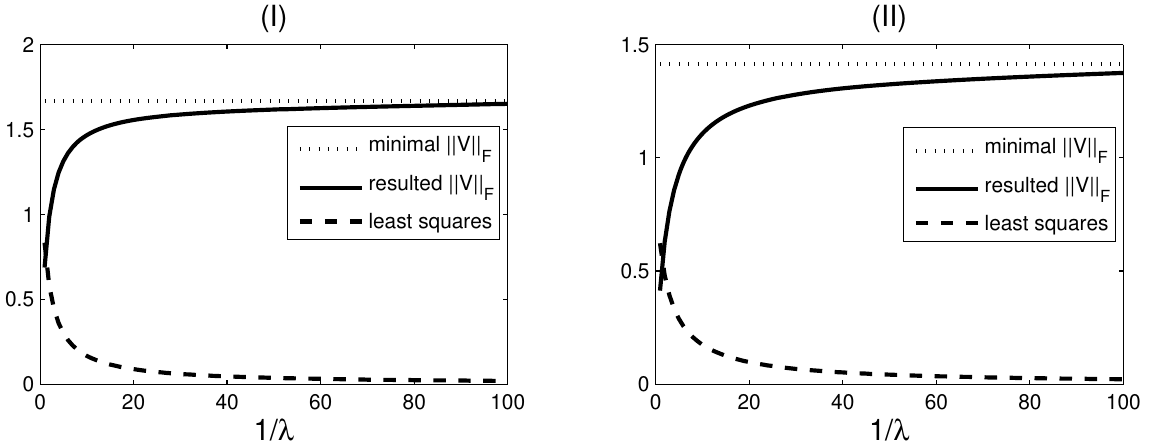}}
\end{center}
\caption{Results of penalized least squares estimation for (I) and (II) in Section \ref{subsec:ufm}.}\label{fig:nonid}
\end{figure}

We use numerical examples to illustrate the penalized least squares method for unidentifiable factor models in Section \ref{subsec:tmo}. Consider the following two covariance matrices,
\begin{eqnarray*}&&\mathrm{(I)}\ \m{\Sigma}=\begin{pmatrix} 2 & 1 \\ 1 & 3 \end{pmatrix}
=\begin{pmatrix} 1/\sqrt{3}\\ \sqrt{3}\end{pmatrix}\begin{pmatrix}1/\sqrt{3}& \sqrt{3}\end{pmatrix}+\begin{pmatrix} 5/3 & 0 \\ 0 & 0 \end{pmatrix},\end{eqnarray*}
\begin{eqnarray*}&&\mathrm{(II)}\ \m{\Sigma}=\begin{pmatrix} 3&-1&-2&2\\-1&2&0&-1\\-2&0&4&-2\\2&-1&-2&2\end{pmatrix}=\begin{pmatrix} 1&-1\\-1&0\\0&2\\1&-1\end{pmatrix}\begin{pmatrix} 1&-1&0&1\\-1&0&2&-1\end{pmatrix}+\begin{pmatrix} 1&0&0&0\\0&1&0&0\\0&0&0&0\\0&0&0&0\end{pmatrix}.\end{eqnarray*}Since all factor models with $(p,q)=(2,1)$ or $(p,q)=(4,2)$ are unidentifiable (Anderson and Rubin 1956), $\m{\Sigma}$ can be written as the form of $\m{A}\m{A}'+\m{V}$ with different $(\m{A},\m{V})$. The above equations show the ones with the smallest $\|\m{V}\|_F$. For (I) and (II), we solve the problem of penalized least squares estimation in \eqref{pmlse} with varying tuning parameter $\lambda$ by Algorithm \ref{ag:cd}, and show the results in Figure \ref{fig:nonid}. It can be seen that, as $1/\lambda$ increases, the resulted $\|\m{V}\|_F$ from the penalized least squares estimates converges to minimal $\|\m{V}\|_F$. At the meantime, the loss of least squares converges to zero, which indicates that the estimates are consistent with the structure of factor models for large $1/\lambda$. Therefore, the proposed penalized least squares estimation can effectively yield the estimates with good dimensionality reduction performance under unidentifiable factor models.

\subsection{Path optimization}\label{subsec:sim}

\begin{figure}[t]\begin{center}
\scalebox{0.7}[0.7]{\includegraphics{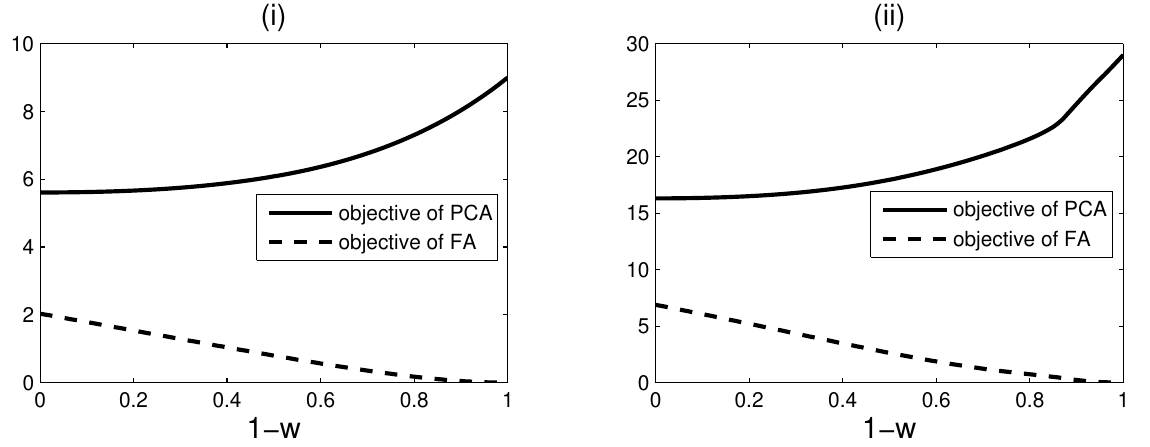}}
\end{center}
\caption{Path from PCA to FA for (i) and (ii) in Section \ref{subsec:sim}.}\label{fig:path}
\end{figure}

We show two examples of the path between PCA and FA introduced in Section \ref{subsec:path}. Consider the following two covariance matrices,
\begin{eqnarray*}&&\mathrm{(i)}\ \m{\Sigma}=\begin{pmatrix} 2 & 1 &1\\ 1 & 3 &1\\1&1&3\end{pmatrix}=\begin{pmatrix} 1 \\ 1\\1\end{pmatrix}\begin{pmatrix} 1 & 1 & 1\end{pmatrix}+\begin{pmatrix} 1 & 0 & 0\\ 0 & 2 & 0\\ 0 & 0 & 2\end{pmatrix},
\\&&\mathrm{(ii)}\ \m{\Sigma}=\begin{pmatrix} 3&-2&-1&1&-2\\-2&5&0&-2&2\\-1&0&4&0&1\\1&-2&0&4&-1\\-2&2&1&-1&5\end{pmatrix}=\begin{pmatrix} 1&-1\\-2&0\\0&1\\1&0\\-1&1\end{pmatrix}\begin{pmatrix} 1&-2&0&1&-1\\-1&0&1&0&1\end{pmatrix}+\begin{pmatrix} 1&0&0&0&0\\0&1&0&0&0\\0&0&3&0&0\\0&0&0&3&0\\0&0&0&0&3\end{pmatrix}.\end{eqnarray*}
We solve the path optimization problem \eqref{pcafa} with objective functions $(F_{1,\mathrm{PCA}}(\m{T}),F_{1,\mathrm{FA}}(\m{T}))=(\|\m{\Sigma}-\m{T}\|_F^2,f_2(\m{\Sigma}-\m{T}))$. For each $w$, we use Algorithm \ref{ag:US} to compute the solution. The results are displayed in Figure \ref{fig:path}. It can be seen that, as $w$ varies from 1 to 0, the objective value of PCA increases, while that of FA decreases to 0. This indicates that the fitted model  becomes closer to the factor model, and reaches it at last.

\subsection{Examples of fitting the principal component factor model}\label{subsec:real}

We now use two real datasets to fit the principal component factor model in Section \ref{subsec:pcfm}. The first dataset comes from 48 Chinese cities, and contains data of three indices of urban development in the year 2008. These indices are as follows:\begin{eqnarray*}
X_1:&&\text{per-capita investment in the fixed assets (thousand RMB)}
\\X_2:&&\text{per-capita net income of rural people (thousand RMB)}
\\X_3:&&\text{birthrate (\textperthousand)}\end{eqnarray*} All data used in this example can be found on the internet (available from the
author).

Here $p=3$, and thus we only need to consider $q=1$ for fitting a factor model.
We first compute the maximum likelihood estimate of the population covariance matrix as
$$\m{\Sigma}=\begin{pmatrix}  82.5524 &4.6990& -5.6177\\ 4.6990 &4.6262& -1.5502\\ -5.6177&-1.5502&4.7571\end{pmatrix}.$$The first principal component is $\v{r}_1=(0.9971, 0.0441, -0.0614)'$. Consequently, we compute the maximum likelihood estimates of the parameters in the principal component factor model by the iterative method in Section \ref{subsec:pcfm}, and obtain the fitted model,
\begin{equation}\label{realeq}
	\left\lbrace \begin{aligned}
		&X_1=8.8792\,U+\varepsilon_1,\\
		&X_2=0.3928\,U+\varepsilon_2,\\
        &X_3=-0.5466\,U+\varepsilon_3,
	\end{aligned}\right.
\end{equation}where $U\sim N(0,1),\ \varepsilon_1\sim N(0,2.0827^2),\ \varepsilon_2\sim N(0,2.0789^2),\ \varepsilon_3\sim N(0,2.0819^2)$ are independent. The factor loading vector in \eqref{realeq} is $\v{a}=(8.8792, 0.3928, -0.5466)'=79.2938\,\v{r}_1$, which indicates that $\v{a}$ has the same direction as the first principal component. We also conduct the likelihood ratio test for testing whether the covariance structure of the principal component factor model holds, and get $p$-value $=0.1538$.

In \eqref{realeq}, the common factor $U$ can be interpreted as the level of development. With a higher level of development, the two economic indicators, $X_1$ and $X_2$, are expected to be high, but the birthrate has a decreasing trend.

The second dataset contains data of air pollution emissions. It is known that air pollution is one of the world's largest health and environmental problems. A feasible method to prevent air pollution is to analyze the sources of air pollution emissions. We now study the relationship among concentrations (unit: $\mathrm{mg}/\mathrm{m}^3$) of five ions in the air. The ions are as follows:
\begin{eqnarray*}
X_1:&&\text{chloridion}\quad(\text{CL}^-)\\ X_2:&&\text{ammonium ion}\quad({\text{NH}_4}^+)\\ X_3:&&\text{calcium ion}\quad(\text{Ca}^{2+})
\\ X_4:&&\text{sodium ion}\quad(\text{Na}^+)\\ X_5:&&\text{potassium ion}\quad(\text{K}^+)\end{eqnarray*}
The data used in this example are sampled from 56 pollution events in 2001-2007 in a Chinese city, which has suffered from serious air pollution.

Consider $q=2$ for fitting a principal component factor model in Section \ref{subsec:pcfm}.
We first compute the maximum likelihood estimate of the population covariance matrix as
$$\m{\Sigma}=\begin{pmatrix}  134.7&848.1&460.4&118.3&62.0\\848.1&7706.9&2941.1&885.2&296.7\\460.4&2941.1&2189.4&576.9&302.9\\118.3&885.2&576.9&264.8&70.4\\62.0&296.7&302.9&70.4&61.9\end{pmatrix}.$$The first two principal components are $\v{r}_1=(-0.1036, -0.9033, -0.3944, -0.1259, -0.0426)'$ and $\v{r}_2=(-0.0662, 0.4193, -0.8006, -0.4053, -0.1207)'$. Consequently, we compute the maximum likelihood estimates of the parameters in the principal component factor model, and obtain the fitted model,
\begin{equation}\label{realeq2}
	\left\lbrace \begin{aligned}
		&X_1=-9.9518\,U_1-2.1146\,U_2+\varepsilon_1,\\
		&X_2=-86.7484\,U_1+13.3896\,U_2+\varepsilon_2,\\
        &X_3=-37.8793\,U_1-25.5632\,U_2+\varepsilon_3,\\
        &X_4=-12.0859\,U_1-12.9416\,U_2+\varepsilon_4,\\
        &X_5=-4.0875\,U_1-3.8551\,U_2+\varepsilon_5,
	\end{aligned}\right.
\end{equation}where $U_1\sim N(0,1),\ U_2\sim N(0,1),\ \varepsilon_1\sim N(0,4.7523^2),\ \varepsilon_2\sim N(0,10.1974^2),\ \varepsilon_3\sim N(0,7.5504^2),\ \varepsilon_4\sim N(0,10.2346^2),\ \varepsilon_5\sim N(0,3.9732^2)$ are independent. The factor loading vectors in \eqref{realeq2} are $9221.9\,\v{r}_1$ and $1019.6\,\v{r}_2$. We also conduct the likelihood ratio test for testing whether the covariance structure of the principal component factor model holds, and get $p$-value $=0.3036$. The common factors $U_1$ and $U_2$ can be interpreted as two sources of air pollution.

\section{Discussion}\label{sec:diss}

In this paper we have proposed a unified way to handle PCA and FA under a general latent variable model. With specific loss functions, the two methods can be viewed as solutions to matrix optimization problems. Along this way, we have discussed a number of topics, including effective algorithms, penalized methods, and some combinations of PCA and FA.

The use of this paper is twofold. First, many students, including myself of the bygone age, feel confused about the connection and difference between PCA and FA when learning the course of multivariate statistical analysis, because the two methods are much different from other contents in textbooks such as statistical inference. The proposed framework to handle PCA and FA is relatively simple, and can complement present textbooks. Hopefully this will help students better understand them.

Second, we have provided new viewpoints to reexamine the two classical methods. New methods for data analysis are presented. As a result researchers can find some future directions from them. For example, there are numerous other loss functions for fitting the model \eqref{lvm}. A natural class of objective functions in \eqref{mo} is $F_1(\m{T})=d\left(N(\v{0},\m{\Sigma}), N(\v{0},\m{T})\right)$, where $d$ evaluates some distance between two distributions. Note that the Wasserstein distance,
which is related to the optimal transport problem, has been widely applied in machine learning and statistics (Panaretos and Zemel 2020). With $d$ being the Wasserstein distance, $F_1$ has a closed form as $F_1(\m{T})={\mathrm{trace}}\left(\m{\Sigma}+\m{T}-2\left(\m{\Sigma}^{1/2}\m{T}\m{\Sigma}^{1/2}\right)^{1/2}\right)$ (Knott and Smith 1984). We conjecture that the solution to \eqref{mo} with this objective function yields PCA. It seems correct even when we remove the constraint $\m{T}\leqslant\m{\Sigma}$ like in \eqref{fmo} after our numerous numerical experiments. This conjecture may construct a connection between PCA and optimal transport. Note that eigenvalue decomposition is computationally expensive with large datasets. We also hope to find efficient iterative algorithms for specific objective functions, which induce PCA, instead of eigenvalue decomposition in PCA (Roweis 1997; Hippert-Ferrer, Bouchard, and Mian et al. 2023).
Another direction is to extend our methods to more complex problems, including supervised PCA (Bair, Hastie, and Paul et al. 2006), sparse PCA (Zou, Hastie, and Tibshirani 2006; Johnstone and Lu 2009), nonlinear dimensionality reduction (Ghojogh, Crowley, and Karray et al. 2023), categorical data analysis (Bhattacharya and Dunson 2012), and estimation of intrinsic dimension (Levina and Bickel 2004).

\section*{Appendix}

\emph{Proof of Theorem \ref{th:F2}}\quad Under Assumption \ref{ass:h}, for fixed $\m{T}$, the solution to \eqref{mo2} with $F_2(\m{T},\m{V})=h(\m{T}+\m{V})+F_1(\m{T})$ is $\m{V}^*(\m{T})=\m{\Sigma}-\m{T}$. Then the optimal $\m{T}^*$ is the solution to \eqref{mo}, and this completes the proof. \qed

\noindent\emph{Proof of Lemma \ref{lemma:afd}}\quad It suffices to show that $\|\m{S}\|_2\geqslant s_{11}$, which is obtained by noting $\|\m{S}\|_2=\max_{\v{x},\ \|\v{x}\|_2=1}\v{x}'\m{S}\v{x}$ and taking $\v{x}=(1,0,\ldots,0)'$. \qed

\noindent\emph{Proof of Theorem \ref{th:pca}}\quad Let $\tilde{\m{T}}=\m{R}'\m{T}\m{R}$. By Assumption \ref{ass:f}, \eqref{mo} is equivalent to
\begin{equation}\label{mom0s}
	\begin{aligned}
		&\min_{\m{T}}\ f(\m{\Lambda}-\tilde{\m{T}}),\\
		&{\text{s.t.}}\ \m{0}\leqslant\tilde{\m{T}}\leqslant\m{\Lambda},\ \mathrm{rank}(\tilde{\m{T}})=q.
	\end{aligned}
\end{equation}
By Assumption \ref{ass:fd}, we only need to consider $\tilde{\m{T}}$ to be a diagonal matrix. By Assumption \ref{ass:fm}, the solution to \eqref{mom0s} is $\tilde{\m{T}}^*=\mathrm{diag}(\lambda_{1},\ldots,\lambda_q, 0,\ldots,0)$, which completes the proof. \qed

\noindent\emph{Proof of Theorem \ref{th:pmle2}}\quad Note that $(\m{T}^*, \m{V}^*)$ is the solution to \begin{equation*}
	\begin{aligned}
		&\min_{\m{T},\m{V}}\ l(\m{T},\m{V}),\\
		&{\text{s.t.}}\ \m{T}\geqslant\m{0},\ \m{V}\geqslant\m{0},\ \mathrm{rank}(\m{T})=q,\ \|\m{V}\|_2\leqslant\kappa,
	\end{aligned}
\end{equation*}where $\kappa\geqslant\|\m{V}^*\|_2$ is a tuning parameter. This completes the proof. \qed

\noindent\emph{Proof of Theorem \ref{th:fa}}\quad Under model \eqref{fm}, $\tilde{f}_0(\m{\Sigma}-\m{T}^*)=\tilde{f}_0(\m{V})\leqslant p$. By \eqref{f0}, for any matrix $\m{S}$ satisfying the constraints in \eqref{mo}, $\tilde{f}_0(\m{\Sigma}-\m{S})<\tilde{f}_0(\m{\Sigma}-\m{T}^*)$ if and only if $\m{\Sigma}-\m{S}=\m{0}$, which is a contradiction since $\mathrm{rank}(\m{S})=q<\mathrm{rank}(\m{\Sigma})$. \qed

\section*{Acknowledgements}

This work is partially supported by National Key R\&D Program of China (Grant nos. 2021YFA1000300, 2021YFA1000301, and 2021YFA1000303) and the National Natural Science Foundation of China (Grant no. 12171462).

\vspace{1cm} \noindent{\Large\bf References}

{\begin{description}

\footnotesize
\item{}
Anderson, T. W. (1984), \textit{An Introduction to Multivariate Statistical Analysis}, Second Edition, John Wiley \& Sons.

\item{}
Anderson, T. and Rubin, H. (1956), Statistical inference in factor analysis. In: Neyman, J., ed., Proc. 3rd Berkeley Symp. Mathematical Statistics and Probability, vol. V, Berkeley, CA: University of California Press, pp. 111-150.

\item{}
Bhattacharya, A. and Dunson, D. B. (2012), Simplex factor models for multivariate unordered categorical data. \textit{Journal of the American Statistical Association},  {\bf 107}, 362-377.

\item{}
Bair, E., Hastie, T., Paul, D., and Tibshirani, R. (2006), Prediction by supervised principal components. \textit{Journal of the American Statistical Association},  {\bf 101}, 119-137.

\item{}
Bartholomew, D., Knott, M., and Moustaki, I. (2011), \textit{Latent Variable Models and Factor Analysis: A Unified Approach}, Third Edition, John Wiley \& Sons.


\item{}
Ghojogh, B., Crowley, M., Karray, F., and Ghodsi, A. (2023), \textit{Elements of Dimensionality Reduction and Manifold Learning}, Springer Nature.

\item{}
Giudici, P. and Stanghelline, E. (2001). Bayesian inference for graphical factor analysis models. \textit{Psychometrika}, {\bf 66}, 577-592.

\item{}
Grzebyk, M., Wild, P., and Chouani\`{e}re, D. (2004), On identification of multi-factor models with correlated residuals. \textit{Biometrika}, {\bf 91}, 141-151.

\item{}
Hippert-Ferrer, A., Bouchard, F., Mian, A., Vayer, T., and Breloy, A. (2023). Learning graphical factor models with riemannian optimization. In Joint European Conference on Machine Learning and Knowledge Discovery in Databases (pp. 349-366). Cham: Springer Nature Switzerland. arXiv:2210.11950v2


\item{}
Levina, E., and Bickel, P. (2004), Maximum likelihood estimation of intrinsic dimension. \textit{Advances in Neural Information Processing Systems}, {\bf 17}. 

\item{}
Johnstone, I. M. and Lu, A. Y. (2009), On consistency and sparsity for principal components analysis in high dimensions. \textit{Journal of the American Statistical Association}, {\bf 104}, 682-693.

\item{}
Jolliffe, I. T. (2002), \textit{Principal Component Analysis}, Second Edition, New York: Springer.

\item{}
Jolliffe, I. T. and Cadima, J. (2016), Principal component analysis: a review and recent developments. \textit{Phil. Trans. R. Soc. A} {\bf 374}: 20150202.

\item{}
Knott, M. and Smith, C. S. (1984), On the optimal mapping of distributions. \textit{Journal of Optimization Theory and Applications},  {\bf 43}: 39-49.

\item{}
Kolda, T. G., Lewis, R. M., and Torczon, V. (2003), Optimization by direct search: New perspectives on some classical and modern methods. \textit{SIAM review}, {\bf 45}: 385-482.

\item{}
Panaretos, V. M. and Zemel, Y. (2020), \textit{An Invitation to Statistics in Wasserstein Space}, Springer.

\item{}
Roweis, S. (1997). EM algorithms for PCA and SPCA. \textit{Neural Inf. Proc. Syst.},  {\bf 10}: 626-632.

\item{}
Shapiro, A. (1985), Identifiability of factor analysis: some results and open problems. \textit{Linear Algebra and Its Applications}, {\bf 70}: 1-7.

\item{}
Srebro, N. and Jaakkola, T. (2003), Weighted low-rank approximations. In ICML, Vol. 3, 720-727.

\item{}
Tipping, M. E. and Bishop, C. M. (1999), Probabilistic principal component analysis. \textit{Journal of the Royal Statistical Society Series B: Statistical Methodology}, {\bf 61}, 611-622.

\item{}
Tseng, P. (2001), Convergence of a block coordinate descent method for nondifferentiable minimization. \textit{Journal of Optimization Theory and Applications}, {\bf 109}, 475-494.

\item{}
Zou, H. Hastie, T., and Tibshirani, R. (2006), Sparse principal component analysis. \textit{Journal of Computational and Graphical Statistics}, {\bf 15}, 265-286.

\end{description}}

\end{document}